%% file: root.tex
\documentclass[letterpaper, 10 pt, conference]{ieeeconf}  

\IEEEoverridecommandlockouts                              

\usepackage{graphics} 
\usepackage{times} 
\usepackage{amsmath} 
\usepackage{amssymb}  
\usepackage{mathtools,amsfonts,amssymb,bm}
\usepackage{hyperref}
\usepackage[demo]{graphicx}
\usepackage{subcaption}
\usepackage{multirow}
\usepackage{xcolor}
\usepackage{soul}
\usepackage{cite}
\usepackage{float}
\usepackage{lipsum}
\usepackage{algorithm}
\usepackage{algpseudocode}

\title{\LARGE \bf
Online Action Recognition for Human Risk Prediction \\ with Anticipated Haptic Alert via Wearables
}

\author{Cheng Guo$^{1,2}$, Lorenzo Rapetti$^{1}$, Kourosh Darvish$^{3}$, Riccardo Grieco$^{1}$, Francesco Draicchio$^{4}$, Daniele Pucci$^{1,2}$
\thanks{*This work was supported by the Italian National Institute for Insurance against Accidents at Work (INAIL) ergoCub Project.}
\thanks{$^{1}$Artificial and Mechanical Intelligence at Istituto Italiano di Tecnologia, Center for Robotics Technologies, Genova, Italy. 
        email: {\tt\small { firstname.lastname@iit.it}}}%
\thanks{$^{2}$School of Computer Science, The University of Manchester, Manchester, United Kingdom.}%
\thanks{$^{3}$University of Toronto, Toronto, Ontario, Canada.}%
\thanks{$^{4}$Department of Occupational and Environmental Medicine, Epidemiology and Hygiene, INAIL, Rome, Italy.}%
}

\begin{document}

\maketitle
\thispagestyle{empty}
\pagestyle{empty}

\begin{abstract}

\input{sections/00_abstract}

\end{abstract}

\input{sections/01_introduction}
\input{sections/02_background}
\input{sections/03_methods}

\input{sections/04_validation}

\input{sections/05_conclusions}






\bibliographystyle{IEEEtran}
\bibliography{refs}

\end{document}

%% file: sections/00_abstract.tex
This paper proposes a framework that combines online human state estimation, action recognition and motion prediction to enable early assessment and prevention of worker biomechanical risk during lifting tasks. The framework leverages the NIOSH index to perform online risk assessment, thus fitting real-time applications. In particular, the human state is retrieved via inverse kinematics/dynamics algorithms from wearable sensor data. Human action recognition and motion prediction are achieved by implementing an LSTM-based \emph{Guided Mixture of Experts} architecture, which is trained offline and inferred online. With the recognized actions, a single lifting activity is divided into a series of continuous movements and the \emph{Revised NIOSH Lifting Equation} can be applied for risk assessment. Moreover, the predicted motions enable anticipation of future risks. A haptic actuator, embedded in the wearable system, can alert the subject of potential risk, 
acting as an active prevention device. The performance of the proposed framework is validated by executing real lifting tasks, while the subject is equipped with the iFeel wearable system. The source code for this paper is available at \href{https://github.com/ami-iit/paper_guo_2023_humanoids_lifting_risk_prediction}{$https://github.com/ami-iit/paper \textunderscore guo \textunderscore 2023 \textunderscore humanoids \textunderscore lifting \textunderscore risk \textunderscore prediction$}.


%% file: sections/01_introduction.tex
\section{INTRODUCTION}
\label{sec:introduction}
Work-related low-back disorders (WLBDs) still represent a societal challenge that 
threat the health conditions of working adults~\cite{Health1981}. Among the large variety of their causes, 
payload lifting tasks in industrial environments play a pivotal role in determining poor ergonomic conditions that favor WLBDs \cite{Kuijer2014, Lu2014, Waters2011}. In this context, ergonomics techniques to assess the quality of work conditions emerged, albeit based on qualitative questioners that are often costly and inconvenient to apply 
for dynamically changing work environments. It is then essential to develop quantitative scalable systems that online monitor human ergonomics and that potentially alert the worker before endangering health conditions.  
This paper proposes a framework that combines wearable sensors and haptic devices, learning-based prediction algorithms and traditional lifting ergonomics to enable early assessment and active prevention of worker biomechanical risk during lifting task execution. 
\begin{figure}[H]
    \centering
    \includegraphics[scale=0.21]{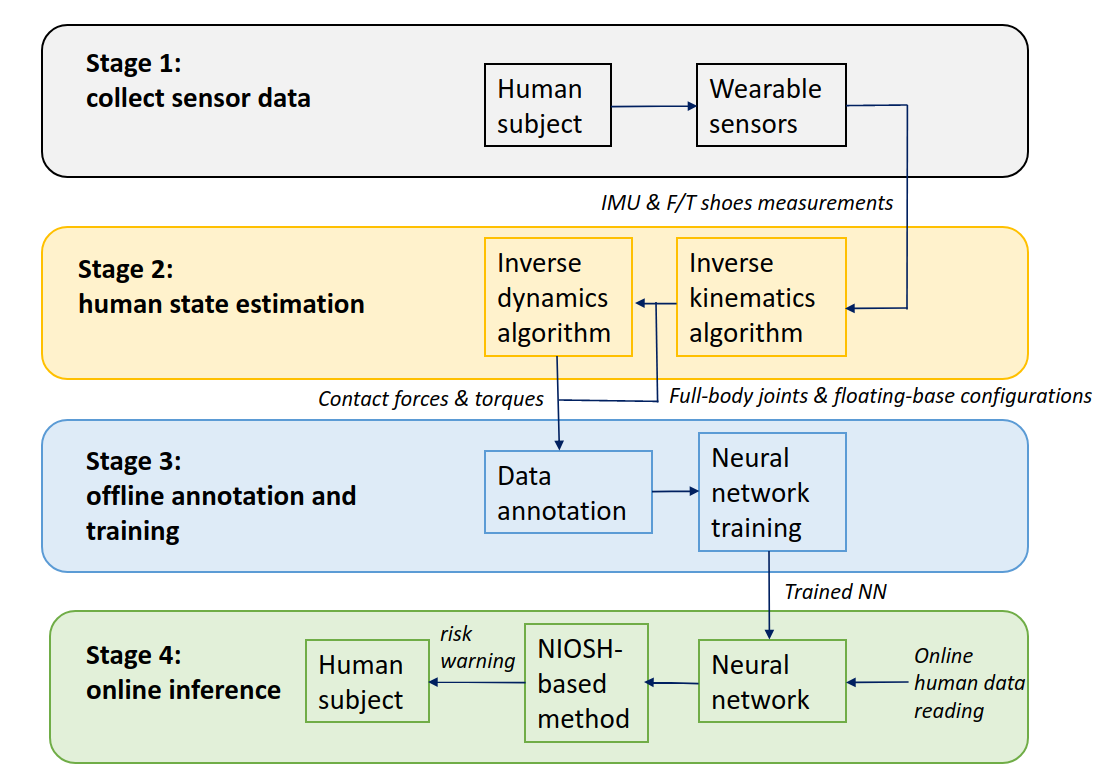}
    \caption{An overview of the proposed framework.}
    \label{fig:general_framework}
\end{figure}

The \emph{Revised NIOSH Lifting Equation} (RNLE) is a renowned 
tool for assessing two-handed manual lifting ergonomics -- it is published by the National Institute for Occupational Safety and Health (NIOSH) \cite{Waters1993, Waters1994}. The RNLE defines a \emph{Recommended Weight Limit} (RWL) and a \emph{Lifting Index} (LI) based on payload weight, which may lead to reliable risk assessment for  
WLMDs \cite{Waters2011}. Unfortunately, approximately 35$\%$ of lifting tasks and 63$\%$ of workers can not be assessed by means of the RNLE due to its limited number of parameters and system constraints \cite{Dempsey2002}. To overcome such limitations, further approaches have been proposed to assess lifting-related risks, e.g., \emph{L5-S1 Internal Forces} \cite{Lavender2003}, \emph{Mechanical Energy Consumption} \cite{Ranavolo2017} and \emph{Muscles Co-Activation} \cite{Ranavolo2015}. However, these offline ergonomics evaluation tools are not flexible enough to be used directly in an unstructured work environment. 

As an attempt towards online human ergonomics evaluation, observational methods -- like the \emph{Rapid Entire Body Assessment (REBA)} and \emph{Rapid Upper Limb Assessment (RULA)} -- are leveraged for human-robot interaction \cite{Shafti2019}.
The human data are measured by wearable sensors and an estimation of motion's ergonomics is provided by automatically fulfilling the worksheet. More recently, real-time tools for tracking joint compressive forces during robot interactions are employed \cite{Fortini2020ROMAN}. 
Analogously, the overloading joint torques can be computed using the displacement of the center of pressure during heavy lifting tasks, returning visual feedback of the worker state \cite{fortini2020framework}.
%
For manual lifting tasks, the existing attempts are either overly generic, e.g.  \cite{Shafti2019}, or limited by hardware settings, e.g. portability restriction \cite{fortini2020framework}. They also lack the ability to alert the worker in advance, beforehand that biomechanical risks endanger health conditions. 

Generally, human action recognition and motion prediction are tackled as two separate issues.
Action recognition can be addressed as a classification problem, solved by applying supervised learning methods \cite{Zhao2019, Ji2012}. Motion prediction is more often regarded as a regression problem, that has been addressed for example by means of generative adversarial networks \cite{Hernandez2019}, graph convolutional networks \cite{Mao2019}, dropout auto-encoder LSTM \cite{Ghosh2017}. 
In \cite{Kourosh2022}, it was proposed the \emph{Guided Mixture of Experts} (GMoE) framework that can resolve these two problems simultaneously, having the potential to simplify the architecture for motion prediction and risk assessment.

This paper proposes a learning-based approach that enables predictions of worker biomechanical risk during lifting tasks with anticipated haptic feedback. We use IMU-based sensing systems that show some advantages over vision-based approaches when used for human motion tracking due to easier calibration and a more convenient application in wider, partially occluded spaces. Moreover, the wearable device can integrate the actuation unit to provide feedback to the subject.
The employed wearable system is composed of 10 IMUs with haptic actuators and a pair of Force/Torque shoes.
The contribution of this paper is threefold. First, we develop a system that can monitor human ergonomics online in the context of lifting activities. To do so, we propose a framework that integrates both the human state estimation algorithm and human action/motion prediction method, enabling the RNLE to not only estimate but also predict lifting risk continuously. Second, we adapted the GMoE \cite{Kourosh2022} approach for recognizing a set of predefined actions that compose a complete lifting activity. The GMoE network is trained on a data set collected in a laboratory environment. Finally, we validate online the proposed framework via an experimental analysis conducted on lifting tasks.

The paper is organized as follows. In Section \ref{sec:background} we introduce the underlying technologies used in our research. In Section \ref{sec:methods} the proposed framework is clarified, including the implementation of RNLE-based human lifting ergonomics monitoring system. Section \ref{sec:validation} presents an experimental analysis conducted on a human subject. At last, Section \ref{sec:conclusions} concludes the paper.

%% file: sections/02_background.tex
\section{BACKGROUND}
\label{sec:background}
\subsection{Wearable System}
Sensing technologies are used to collect inputs from the environment by measuring physical quantities. In this research, we employed \emph{iFeel}\footnote{\href{https://ifeeltech.eu/}{$https://ifeeltech.eu/$}}, a wearable sensors system developed at Istituto Italiano di Tecnologia (IIT) to monitor human states and provide responses \cite{sortino2023}. The system integrates both motion capture and force/torque sensing. Motion capture aims at tracking and recording the motion, based on IMU sensors. IMUs ensure high-frequency data and low latency, making \emph{iFeel} suitable for real-time motion tracking. F/T sensors are used for measuring and regulating contact forces/torques when interacting with the environment. 

\subsection{Human Modeling and State Estimation}
\label{sec:human_modeling_estimation}
The human is modeled as a floating-base multi-rigid-body dynamic system \cite{Latella2019}. The system configuration is represented by $q = (q_b, s)$, where $q_b$ implies the floating-base pose (position and orientation) w.r.t. the inertial frame $\mathcal{I}$ and $s$ is the joint position vector. The system velocity and acceleration are denoted by ${\nu}$ and $\dot{\nu}$ respectively. The n+6 equations describing human motion with $n_c$ applied external wrenches is \cite{Kourosh2022}:
\begin{equation}\label{human_dynamics}
    M(q)\dot{\nu} + C(q, \nu)\nu + g(q) = B\tau + \displaystyle\sum_{k=1}^{n_c} {J_{k}^{T}(q) f_{k}^{c}},
\end{equation}
where $M(q)$ and $C(q, v)$ represent respectively the mass and Coriolis effect matrix. $g(q)$ is the vector of the gravitational term. $B$ is a selector matrix for joint torques $\tau$. $J_{k}$ is the \emph{Jacobian} mapping the system velocity with the \emph{k-th} link velocity that is associated with the external wrench $f_{k}^{c}$. $n$ indicates the number of joints.

To estimate in real time the system configuration $q$ and its velocity $\nu$, a \emph{dynamical inverse kinematics optimization} approach is proposed in \cite{Rapetti2020}. The idea is to minimize the distance between the computed state configuration $(q(t), \nu(t))$ with the target measurements. First, the measured velocity is corrected using a rotation matrix. Then, to compute the state velocity, the constrained inverse differential kinematics for the corrected velocity vector is solved as a QP optimization problem. At last, the state velocity is integrated to obtain the configuration $q(t)$. For the base estimation, force/torque measurements are applied to determine the location of contacts. Then base estimation can be solved as part of the \emph{dynamical inverse kinematics framework} \cite{Ramadoss2022}.

In \cite{Latella2019}, the estimation of the human dynamics is performed by means of a Maximum-A-Posteriori (MAP) algorithm. The overall system dynamics can be reshaped to an equivalent compact matrix form. In this (Gaussian) domain, the vector of human kinematics/dynamics quantities can be regarded as stochastic variables. Given the measurement reliability, the solution is computed by maximizing the probability of this kinematics/dynamics vector.  

\subsection{Guided Mixture of Experts}
The problem of simultaneous human action recognition and motion prediction is solved jointly by GMoE,  a learning-based approach proposed in \cite{Kourosh2022}. Given the past human states $x_{k-i}$, external forces $f_{k-i}^c$ and hidden states $r_{k-i}$, the next optimal human state $x_{k+1}^{*}$ can be formulated as:
\begin{equation}\label{mapping_func}
    x_{k+1}^{*} = \mathcal{H}^{*}(x_k,..., x_{k-N}, f_k^c,..., f_{k-N}^c, r_k,..., r_{k-N}),
\end{equation}
where the optimal mapping $\mathcal{H}^{*}$ is learned from human demonstration. By recursively applying equation (\ref{mapping_func}), we can predict the future human states for the time horizon T.

In terms of $r_{k-i}$, we only consider human symbolic actions as the hidden states for simplification and estimate it as a classification problem. Hence, equation (\ref{mapping_func}) can be further rearranged as:
\begin{subequations}
\begin{align} \label{action_prediction}
&\tilde{a}_{k+1} = \mathcal{D}_1^*(x_k, ..., x_{k-N}, f_k^c, ..., f_{k-N}^c) ~,\\
\label{motion_prediction}
&\tilde{x}_{k+1}, \tilde{f}_{k+1}^c  = \mathcal{D}_2^*(x_k, ..., x_{k-N}, f_k^c, ..., f_{k-N}^c, \tilde{a}_{k+1}) ~.
\end{align}
\end{subequations}
where $\tilde{a}_{k+1}$ denotes the estimated human next action, $\mathcal{D}_1^*$ and $\mathcal{D}_2^*$ are two optimal mappings to learn.

Integrating the idea of Mixture of Experts (MoE), the \emph{gating network} is guided to learn mapping $\mathcal{D}_1^*$ as a classification problem for recognizing human actions, while each \emph{expert network} learns $\mathcal{D}_2^*$ as a regression problem to predict human motions associated with each specific action.

\subsection{Revised NIOSH Lifting Equation}
\label{sec:RNLE}
The \emph{Revised NIOSH Lifting Equation} (RNLE) consists of the following two empirical equations:
\begin{subequations}
\begin{align} \label{RWL_eqa}
&\text{RWL} = \text{LC}\cdot\text{HM}\cdot\text{VM}\cdot\text{DM}\cdot\text{AM}\cdot\text{FM}\cdot\text{CM} ~,\\
\label{RI_eqa}
&\text{LI} = \text{$W_{payload}$} / \text{RWL} ~.
\end{align}
\end{subequations}

Equation (\ref{RWL_eqa}) determines a \emph{Recommended Weight Limit} (RWL) for a specific task. Each factor in the equation is either from a qualitative assessment or from geometrical measurements weighted by a multiplier. More precisely, \emph{LC} is the load constant (23kg), \emph{HM} is the horizontal multiplier, \emph{VM} is the vertical multiplier, \emph{DM} is the vertical traveling distance multiplier, \emph{AM} is the asymmetry multiplier, \emph{FM} is the frequency multiplier and \emph{CM} is the coupling multiplier. 

The \emph{Lifting Index} (LI) provides an estimate of the physical stress level, which is obtained in equation (\ref{RI_eqa}) by dividing the payload weight \emph{$W_{payload}$} by the recommended weight limit. A LI smaller than 1.0 implies a safe condition for working healthy employees, while a higher value of LI denotes an increasing risk of work-related injuries. 

%% file: sections/03_methods.tex
\section{PROPOSED FRAMEWORK}
\label{sec:methods}
\begin{figure}[htp]
   \centering
   \includegraphics[scale=0.18]{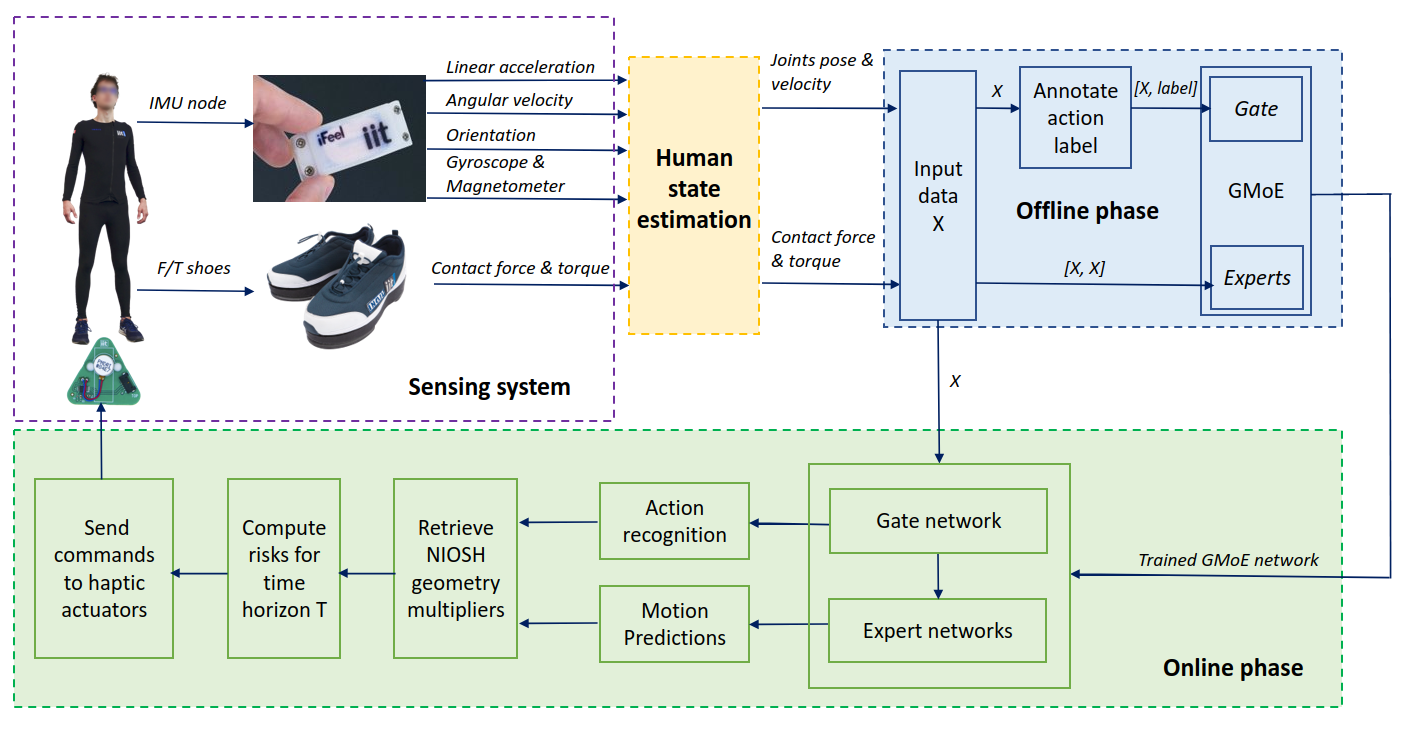}
   \caption{Data flow of the proposed framework, composed of an online and offline phase.}
   \label{fig:pipelne_overview}
\end{figure}

In this research, we propose a four-stage framework, as illustrated in Figure \ref{fig:general_framework}, that integrates methods introduced in Section \ref{sec:methods} for continuously estimating lifting risks and monitoring human ergonomics in real-time applications. A more detailed data flow of working pipelines is shown in Figure \ref{fig:pipelne_overview}.

Firstly, human kinematic measurements and ground contact force/torque are collected by \emph{iFeel node} and \emph{F/T shoes}. The sensor data are then regarded as the \emph{targets} for estimating human full-body joints/floating-base configurations (e.g. positions and velocities) and external feet wrenches (e.g. forces and torques) via Inverse Kinematics (IK) and Inverse Dynamics (ID) algorithms. Afterwards, the output of the \emph{human state estimation} module is manually annotated according to pre-defined action labels for the GMoE network training. Finally, during the online phase, combining the outputs of GMoE and IK/ID modules, the \emph{NIOSH-based method} module is able to provide risk predictions for a given time horizon and thus send commands to haptic actuators worn by human subject.

\subsection{Data Preparation}
\label{subsec:make_data}
To apply GMoE for a lifting task scenario, we build a 15-minute dataset, with data sampled at a frequency of 100Hz. The dataset consists of two volunteers executing three types of lifting tasks repetitively, lasting for 150 seconds each. The volunteer is asked to naturally lift a 3kg payload to a certain height without twisting the upper trunk. The lifting height ranges from 68cm to 92cm, while the other variables (e.g., horizontal distance, payload weight, etc.) remain the same. 

\begin{figure}[htp]
     \centering
     \begin{subfigure}[b]{0.15\textwidth}
         \centering
         \includegraphics[width=\textwidth]{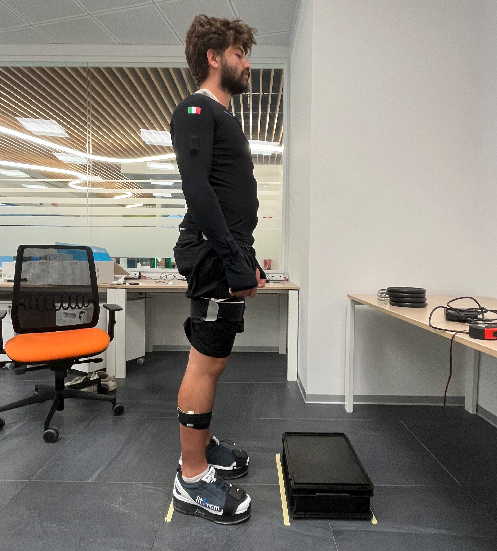}
         \caption{\emph{standing}}
         \label{fig:lift_standing}
     \end{subfigure}
     \hfill
     \begin{subfigure}[b]{0.15\textwidth}
         \centering
         \includegraphics[width=\textwidth]{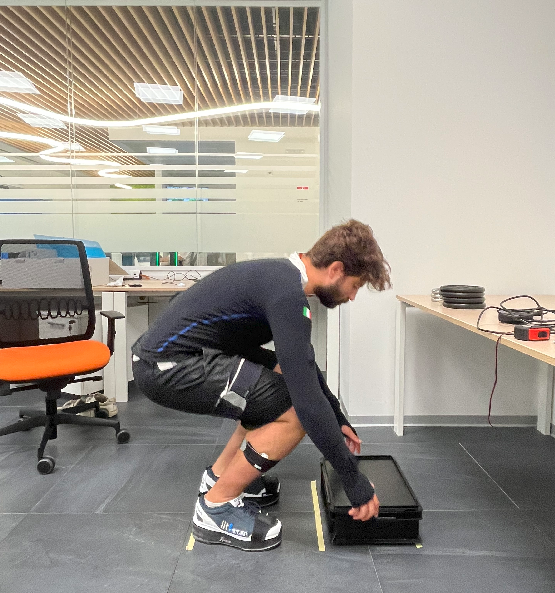}
         \caption{\emph{squatting}}
         \label{fig:lift_squatting}
     \end{subfigure}
     \hfill
     \begin{subfigure}[b]{0.15\textwidth}
         \centering
         \includegraphics[width=\textwidth]{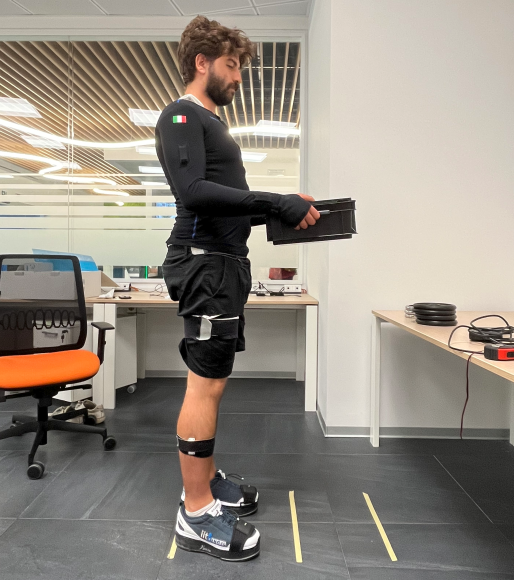}
         \caption{\emph{rising}}
         \label{fig:lift_rising}
     \end{subfigure}
        \caption{The three phases composing the lifting activity.}
        \label{fig:lift_activity}
\end{figure}

Assume the human subject starts with a \emph{standing} pose, a natural sequence of actions during a single lifting activity consists of \emph{squatting}, \emph{rising} and back to \emph{standing} pose again. The lifting risks are most likely to happen during \emph{squatting} and \emph{rising} phases. To apply the NIOSH equation, we must identify the starting and ending moments of each action to establish the initial and final positions of the human subject. For this purpose, we segment a single lifting activity into three continuous phases, each corresponding to a specific action, as illustrated in Figure \ref{fig:lift_activity}.

Given the high cost associated with manual labeling, we have developed an autonomous tool aimed at enhancing the efficiency of annotation. In this labeling process, the estimated data for the entire human body is visualized using a URDF model, while data are streamed in a terminal with a fixed frequency. By observing the action change of the URDF model, an action label is carefully assigned to the current data frame. As long as no new label is assigned by the user, the following data frames are considered to belong to the previous action. More precisely, the transition between \emph{standing} and \emph{squatting} is discerned by observing the bending of the knee. Once the \emph{squatting} action is reaching the end, the ascent of the pelvis denote the beginning of the \emph{rising} phase. The accomplishment of \emph{rising} is detected when, observing a totally erect trunk, the label is assigned as \emph{standing} once again. In the end, the annotated data are divided into three subsets, 70$\%$ for training, 20$\%$ for validation, and the last 10$\%$ for test.

\subsection{GMoE for Lifting Activity}
\begin{figure}[t]
    \centering
    \includegraphics[scale=0.24]{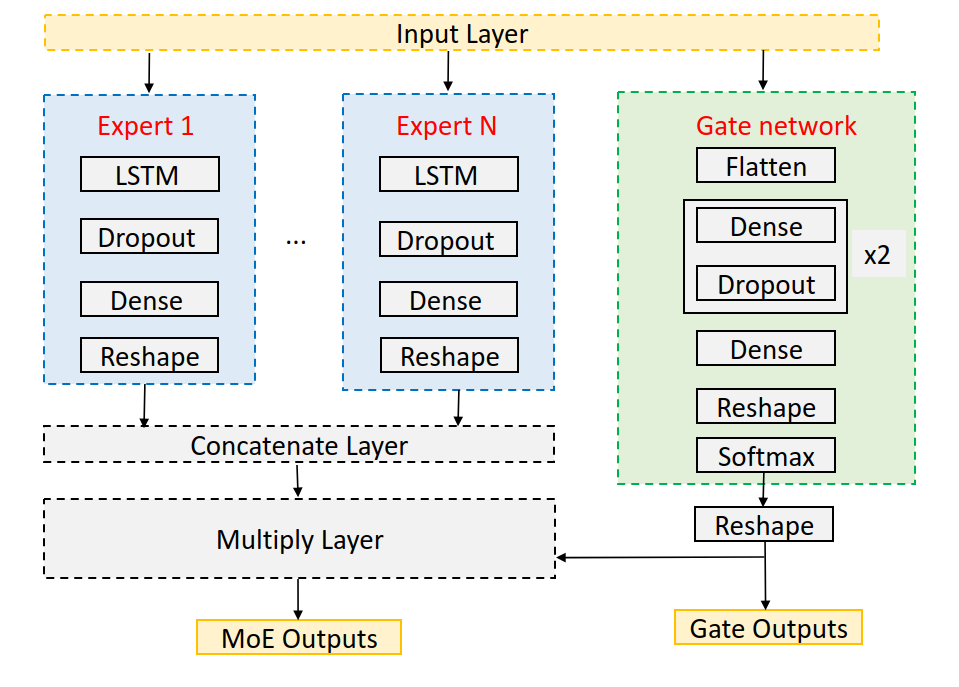}
    \caption{Adapted structure of Guided Mixture of Experts architecture for action recognition and motion prediction.}
    \label{fig:moe_model}
\end{figure}

To achieve simultaneous action recognition and motion prediction, we adopt the network model proposed in \cite{Kourosh2022}. Since three actions are considered in our case, the implemented GMoE architecture is composed of three \emph{expert} networks and one \emph{gate} network, as illustrated in Figure \ref{fig:moe_model}.

The input layer is of size 10x74, where 10 represents the window size for reading past data frames, while 74 is the number of input features, consisting of 31 joint positions, 31 velocities, and 12 contact forces/torques. The \emph{gate} network output layer has size 3x50, where 3 denotes the action categories, and 50 denotes the number of future frames for which action probabilities are computed. It should be noted that when generating time series data, we practically take one data frame every three time steps, such that the period between two adjacent data frames in an input sequence is 30ms. Therefore the total prediction time horizon is 1.5 seconds. Similarly, the output size of each \emph{expert} network is 3x50x43, where 31 joints' positions and 12 foot wrenches are considered (in total 43 output features), excluding the joints' velocities.

During the training phase, the loss function $L_1$ associated with \emph{gate} network and loss function $L_2$ associated with \emph{expert} network are chosen as categorical cross-entropy loss and mean squared error loss, respectively. The total loss function L for GMoE is expressed as a linear combination of $L_1$ and $L_2$:
\begin{equation} \label{loss}
\begin{split}
L & = b_1L_1 + b_2L_2 \\
 & = -\frac{b_1}{2M}\sum_{t=1}^{T}\sum_{j=1}^{M}\sum_{i=1}^{N} a_i^{j,t}log(\tilde{a}_i^{j,t}) \\
 & + \frac{b_2}{2M}\sum_{t=1}^{T}\sum_{j=1}^{M}\|\sum_{i=1}^{N}\tilde{a}_i^{j,t}\tilde{\bm{y}}_i^{j,t}-\tilde{\bm{y}}^{j,t}\|_2
\end{split}
\end{equation}
where $b_1$ and $b_2$ are manually chosen for the convergence of both classification and regression problems (in this case, $b_1$ is 1.0 and $b_2$ is 0.5 for faster convergence of \emph{gate} network), T is the prediction time horizon, M is the total number of data frames, N is the number of experts, scalar value $a_i^{j,t}$ and vector $\tilde{\bm{y}}_i^{j,t}$ denote for human action and motion ground truth associated with \emph{i}-th action and \emph{j}-th data frame at time instance t in the future, operator $\tilde{\cdot}$ represents prediction values of both action recognition and future motions. To update the network weights, \emph{Adam} optimizer is applied with epsilon equals 1e-6. Moreover, early stopping technique and adaptive learning rate are used to avoid overfitting or local optimum. 

\subsection{Risk Prediction and Haptic Alert}
\label{sec:RNLE_implementation}
As mentioned in Section \ref{subsec:make_data}, action recognition is used to determine the origin and destination time point of each action during a single lifting activity. Once the origin status is identified, each following instant can be considered a temporary destination status, which makes it possible to use NIOSH equation to compute risk at that time. Until next action is detected, the NIOSH equation can be applied repeatedly without violating any constraints. Furthermore, by making use of predicted motions, we are also capable of predicting potential risks in the future for a given time horizon. The process of estimating and predicting risks is demonstrated in Algorithm \ref{algorithm:risk_prediction}. Once any potential risk is detected, a command will be sent to the haptic actuator mounted on the human's back. The vibration intensity of the actuator corresponds to the predicted risk level. The human can thus take appropriate measures based on the vibrotactile feedback, i.e., to abort the task immediately or adjust only the lifting posture.
\begin{algorithm}[b]
\caption{Risk prediction using RNLE}
\label{algorithm:risk_prediction}
\begin{algorithmic}
\Require action at $t_0$: $A_{t_0}$, action at $t$: $A_t$, motion prediction at $t$ for future N steps: $M_{t}^{t+N}$, human origin status at $t_0$: $S_{t_0}$, NIOSH variables: $A$, $C$, $F$
\Ensure risk prediction at $t$ for future N steps: $R_t^{t+N}$
\State Initialize $R_t^{t+N}$
\While{$True$}
\If{$A_t$ is not $A_{t_0}$} \Comment{Detect next action}
    \State $A_{t_0} \gets A_t$
    \State $S_{t_0} \gets getHumanStatus(M_{t}^{t+N}[0])$
\EndIf
\For{each item $i$ in $M_t^{t+N}$}
    \State $S_t \gets getHumanStatus(M_{t}^{t+N}[i]))$ 
    \State $H, V, D \gets getVariables(S_{t_0}, S_t)$
    \State $R_t^{t+N}.append(RNLE(H,V,D,A,C,F))$
\EndFor
\State return $R_t^{t+N}$
\EndWhile
\end{algorithmic}
\end{algorithm}

In practice, the action transition cost about 0.5s, which affects the accuracy of action detection. To retrieve more precise NIOSH variables, we implement an approach to compensate action change delay. At each moment, when the probability of previously recognized action is growing, the current action label maintains the same. Once the probability decreases over a pre-defined threshold, we consider the action transition already starts. Then we search for the action label whose probability increases also over a threshold. 

As shown in Algorithm \ref{algorithm:risk_prediction}, from predicted motions we can update the human model in simulator and retrieve geometry values to compute NIOSH variables \emph{H}, \emph{V} and \emph{D}. Assume that the middle point of human hands is always overlapped with the Center of Mass (CoM) of the payload, \emph{H} can be thus represented as the horizontal distance between the position of the CoM of human hand w.r.t. the frame attached to human foot, while \emph{V} is computed by using the vertical position of the human hand w.r.t. the human foot:
\begin{subequations}
\begin{align} \label{eqa_H}
& H = \frac{H_{LeftHand}^{LeftFoot} + H_{RightHand}^{RightFoot}}{2} ~,\\
\label{eqa_V}
& V = \frac{V_{LeftHand}^{LeftFoot} + V_{RightHand}^{RightFoot}}{2} ~.
\end{align}
\end{subequations}
and vertical traveling distance is denoted as \(D = V_t - V_{t_0}\),  where $V_t$  and $V_{t_0}$ represent the vertical distance at the destination and origin moment, respectively. For simplification, asymmetry angle $A$ is not considered in our case, hence $AM$ constantly equals 1. Lifting frequency is computed as the average number of lifts per minute over a 15-minute period. The coupling situation is considered as \emph{Fair}.


%% file: sections/04_validation.tex
\section{VALIDATION}
\label{sec:validation}
\subsection{Experimental Setup}
To validate the performance of the proposed framework for assessing lifting risk in a real-time application, an experimental analysis is performed in a laboratory environment. A healthy volunteer is asked to perform three different lifting tasks corresponding to varied risk levels. In this setup, the participant’s kinematics state is collected using \emph{iFeel}, which is composed of a set of \emph{iFeel-Nodes} (including sensors and actuators) and a central processing unit \emph{iFeel-Station} (a micro-controlled board). The system operates for whole-body motion tracking via \emph{iFeel-Nodes} that are mounted in pre-defined locations of the \emph{iFeel-Suit}. Each \emph{iFeel-Node} contains a 9-DoF IMU that provides absolute orientation and sensor-based velocity fusion data at a rate of 100 Hz. Once detecting any possible risks, a signal is sent to the haptic actuator of the \emph{ifeel node} mounted on the human waist. The ground reaction forces and torques are retrieved using \emph{iFeel-Shoes} equipped with F/T sensors integrated in the front and rear parts. The collected human data are streamed and resampled via YARP middleware \cite{Metta2006} at a rate of 100Hz. Moreover, as mentioned in section \ref{sec:human_modeling_estimation}, human is modeled as a floating-base multi-rigid-body system considering 13 joints (e.g. T9T8, Right shoulder, etc.). The programs run on a 64-bit i7 2.6GHz laptop which is equipped with 32 GB RAM, Intel(R) UHD Graphics and Ubuntu 20.04 LTS. 

The parameters of performed lifting tasks are listed in Table \ref{tab:task_data}. Specifically, asymmetry angle A equals zero (AM = 1.0), coupling quality is \emph{Fair} (CM = 0.95), and lifting frequency is controlled as 7 lifts/min (FM = 0.7). The payload is evenly distributed inside a square box and the weight value is sent to the framework as an external parameter from the user. 
\begin{table}[]
\centering
\caption{Experimental lifting task variables of RNLE.}
\resizebox{\columnwidth}{!}{%
\begin{tabular}{|c|cclccccc|cc|}
\hline
\multirow{3}{*}{\textbf{\begin{tabular}[c]{@{}c@{}}Task\\ type\end{tabular}}} &
  \multicolumn{8}{c|}{\textbf{RNLE variables}} &
  \multicolumn{2}{c|}{\textbf{RNLE results}} \\ \cline{2-11} 
 &
  \begin{tabular}[c]{@{}c@{}}H\_origin\\ (cm)\end{tabular} &
  \multicolumn{2}{c}{\begin{tabular}[c]{@{}c@{}}H\_end\\ (cm)\end{tabular}} &
  \begin{tabular}[c]{@{}c@{}}V\_origin\\ (cm)\end{tabular} &
  \begin{tabular}[c]{@{}c@{}}V\_end\\ (cm)\end{tabular} &
  \begin{tabular}[c]{@{}c@{}}D\_origin\\ (cm)\end{tabular} &
  \multicolumn{1}{c|}{\begin{tabular}[c]{@{}c@{}}D\_end\\ (cm)\end{tabular}} &
  \multirow{2}{*}{\begin{tabular}[c]{@{}c@{}}L\\ (kg)\end{tabular}} &
  \multicolumn{1}{c|}{\begin{tabular}[c]{@{}c@{}}RWL\_origin\\ (kg)\end{tabular}} &
  \multirow{2}{*}{LI} \\ \cline{2-8} \cline{10-10}
 &
  \multicolumn{1}{l}{HM\_origin} &
  \multicolumn{2}{l}{HM\_end} &
  \multicolumn{1}{l}{VM\_origin} &
  \multicolumn{1}{l}{VM\_end} &
  \multicolumn{1}{l}{DM\_origin} &
  \multicolumn{1}{l|}{DM\_end} &
   &
  \multicolumn{1}{c|}{\begin{tabular}[c]{@{}c@{}}RWL\_end\\ (kg)\end{tabular}} &
   \\ \hline
\multirow{2}{*}{Task 1} &
  47 &
  \multicolumn{2}{c}{63} &
  8 &
  68 &
  60 &
  \multicolumn{1}{c|}{60} &
  \multirow{2}{*}{3} &
  \multicolumn{1}{c|}{5.84} &
  0.51 \\
 &
  0.53 &
  \multicolumn{2}{c}{0.40} &
  0.80 &
  0.98 &
  0.90 &
  \multicolumn{1}{c|}{0.90} &
   &
  \multicolumn{1}{c|}{5.40} &
  0.56 \\ \hline
\multirow{2}{*}{Task 2} &
  47 &
  \multicolumn{2}{c}{63} &
  8 &
  80 &
  72 &
  \multicolumn{1}{c|}{72} &
  \multirow{2}{*}{7} &
  \multicolumn{1}{c|}{5.71} &
  1.23 \\
 &
  0.53 &
  \multicolumn{2}{c}{0.40} &
  0.80 &
  0.99 &
  0.88 &
  \multicolumn{1}{c|}{0.88} &
   &
  \multicolumn{1}{c|}{5.33} &
  1.31 \\ \hline
\multicolumn{1}{|l|}{\multirow{2}{*}{Task 3}} &
  47 &
  \multicolumn{2}{c}{63} &
  8 &
  92 &
  83 &
  \multicolumn{1}{c|}{83} &
  \multirow{2}{*}{10} &
  \multicolumn{1}{c|}{5.64} &
  1.77 \\
\multicolumn{1}{|l|}{} &
  0.53 &
  \multicolumn{2}{c}{0.40} &
  0.80 &
  0.95 &
  0.87 &
  \multicolumn{1}{c|}{0.87} &
   &
  \multicolumn{1}{c|}{5.06} &
  1.98 \\ \hline
\end{tabular}
\label{tab:task_data}
}
\end{table}
During the experiment, the participant is asked to repeat each task three times as steady and natural as possible, such that no jerks appear during lifting. The participant should avoid twisting the upper trunk so that the assumption of zero asymmetry angle is fulfilled. Furthermore, the participant is required to hold the box with both hands while his feet are maintained in a fixed position. The lifting activity is executed slowly, hence every single execution can be regarded as independent from the others. 

\subsection{Results and Discussion}
In this section, we first performed a variety of quantitative evaluations of the adapted GMoE model using an additionally collected unseen dataset, which is in a total of 15248 frames. Then we conducted a qualitative analysis based on the results of previously designed online experiments. 
\subsubsection{Quantitative Evaluation of Action Recognition}
\label{sec: analysis action recognition}
\begin{figure}[t]
   \centering
   \includegraphics[scale=0.4,trim={1cm 0 5.3cm 2cm},clip]{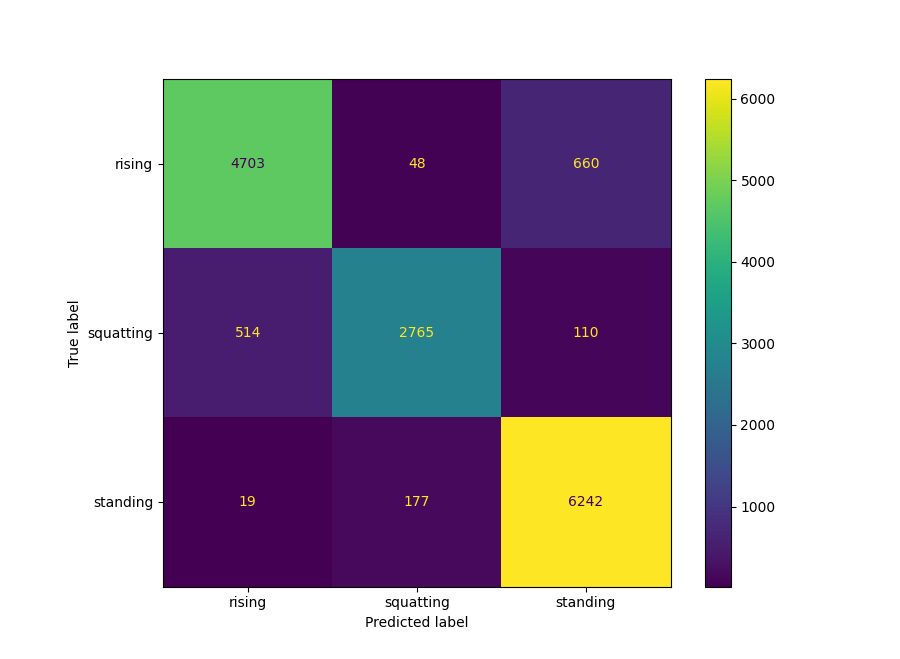}
   \caption{Confusion matrix for the action classification.}
   \label{fig:confusion_matrix}
\end{figure}

In order to assess the action classification performance, a confusion matrix associated with three human lifting actions is presented in Figure \ref{fig:confusion_matrix}.
Based on this confusion matrix, metrics such as \emph{Accuracy}, \emph{Precision}, \emph{Recall} and \emph{F1 score} can be further retrieved. \emph{Accuracy} is the number of correct predictions of all \emph{N} categories divided by the total number of predictions, as shown in Equation \ref{accuracy}, where \emph{total} means the number of all tested samples.
\begin{equation} \label{accuracy}
\begin{split}
Accuracy = \frac{\sum_{i=1}^{N}TP_i}{total}
\end{split}
\end{equation}

\emph{Precision} refers to the proportion of correctly predicted positive instances out of all instances predicted as positive, while \emph{Recall} measures the proportion of correctly predicted positive instances out of all actual positive instances, as shown in Equation \ref{precision} and \ref{recall}, where \emph{i} means each class.
\begin{subequations}
\begin{align} \label{precision}
& Precision = \frac{TP_i}{TP_i + FP_i} ~,\\
\label{recall}
& Recall = \frac{TP_i}{TP_i + FN_i} ~.
\end{align}
\end{subequations}

\emph{F1 score} can be interpreted as a harmonic mean of the \emph{Precision} and \emph{Recall} as shown in Equation \ref{F1_score}.
\begin{equation} \label{F1_score}
\begin{split}
F1 = \frac{2 * Presicion * Recall}{Precision + Recall}
\end{split}
\end{equation}

Table \ref{tab:metrics} summarizes the experimental results of these metrics for each single category classification.
\begin{table}[b]
\centering
\caption{Performance metrics for assessing GMoE model recognizing multi-class human lifting actions.}
\resizebox{\columnwidth}{!}{%
\begin{tabular}{|c|c|c|c|c|}
\hline
\multicolumn{1}{|l|}{} & \textbf{Accuracy} & \textbf{Precision} & \textbf{Recall} & \textbf{F1 score} \\ \hline
\textit{standing}  & \multirow{3}{*}{/} & 0.890 & 0.969 & 0.928 \\ \cline{1-1} \cline{3-5} 
\textit{rising}    &                    & 0.898 & 0.869 & 0.883 \\ \cline{1-1} \cline{3-5} 
\textit{squatting} &                    & 0.925 & 0.816 & 0.867 \\ \hline
\textit{average}   & 0.899              & 0.904 & 0.885 & 0.893 \\ \hline
\end{tabular}
\label{tab:metrics}
}
\end{table}
As we can see, \emph{squatting} has the highest accuracy of 0.925, which indicates that the model has a low rate of falsely labeling instances as this action. On the contrary, \emph{standing} has a relatively low accuracy. This is mainly because the transition period between \emph{rising} and \emph{standing} can be quite ambiguous (also partly due to the fact that the annotated border depends on human judgment), such that it can be hard for the model to distinguish these two phases exactly. Furthermore, both \emph{squatting} and \emph{rising} have relatively lower \emph{Recall} values than \emph{standing}. As explained before, the ambiguity between \emph{rising} and \emph{standing} leads to some false labeling of \emph{standing} when they are actually \emph{rising}. Also, the similarity between the motion patterns of \emph{squatting} and \emph{rising} (they are basically reversed) results in the confusion of them.

\subsubsection{Quantitative Evaluation of Motion Prediction}
\label{sec: analysis motion prediction}
\begin{figure}[t]
   \centering
   \includegraphics[width=1.1\columnwidth]{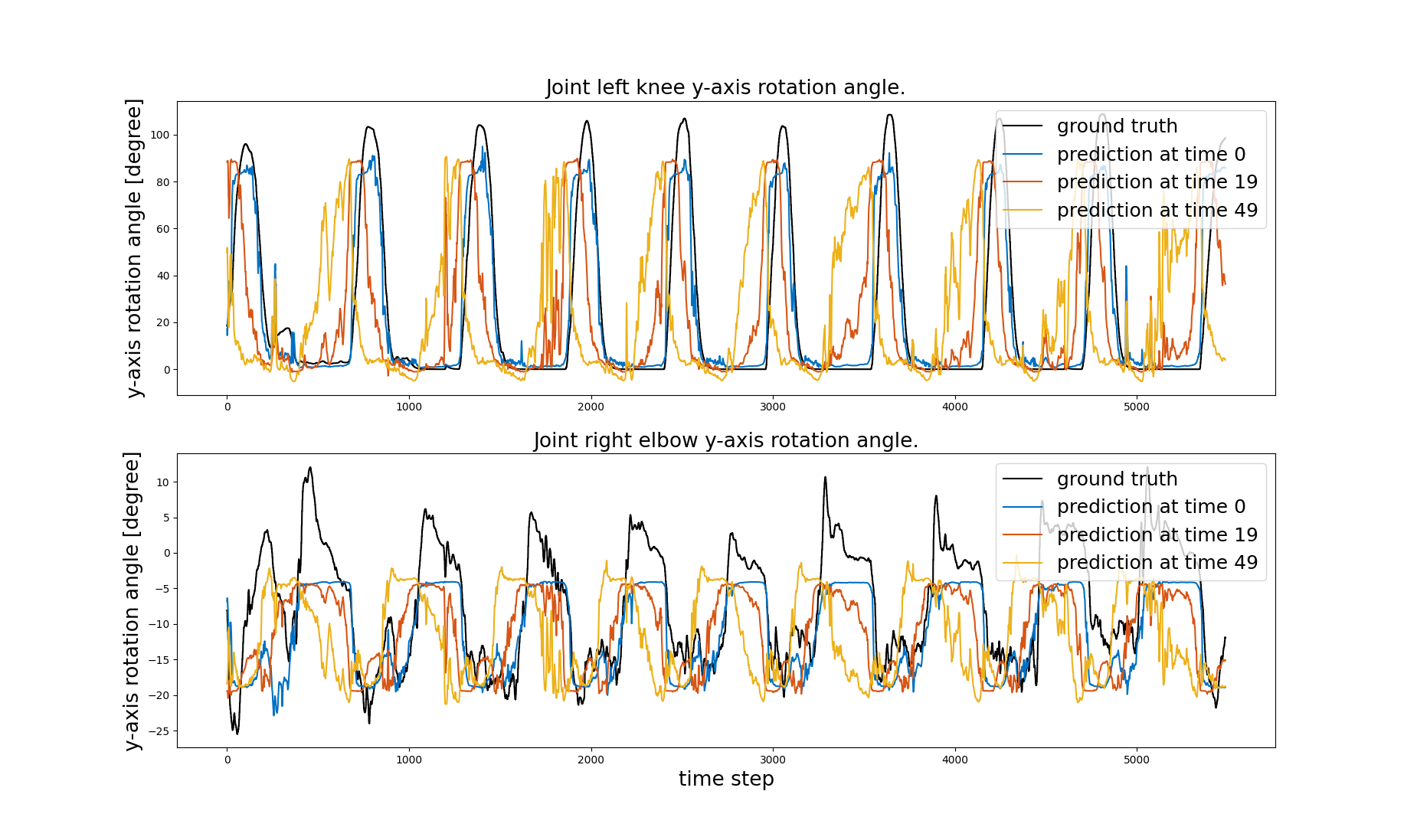}
   \caption{Multi-time-step predictions of the y-axis rotation angle of the joint left knee and right elbow.}
   \label{fig:motion_predictions}
\end{figure}

In the following, we report the performance of GMoE regarding the task of motion prediction. Two key joints (i.e., left knee and right elbow) that can reflect respectively the human upper-/lowerbody motion patterns during a lifting task are chosen. The rotational angles around the y-axis of these two joints during a period of about 5500 frames are demonstrated in Figure \ref{fig:motion_predictions}. The ground truths are depicted in black curves, while the predicted angles at the future time steps 0, 19 and 49 are shown respectively in blue, orange and yellow.

It can be observed from both rows in Figure \ref{fig:motion_predictions} that the predicted y-axis rotational angle at future time step 0 (blue curves) basically captures the motion pattern of the ground truths (black curves), despite the amplitude gaps at peaks. The amplitude differences at peak positions can be more easily observed for the right elbow joint. The predicted rotational angles at future time steps 19 and 49 exhibit a leading phase compared with the ground truth, where the phase differences should match the corresponding prediction time steps. It should be noted that the predictions at future time step 49 suffer more from uncertainties, which is reflected by the frequently appearing sharp fluctuations. This may be due to the fact that the model only has very limited historical information, yet to make a further prediction in the future, it is apparently insufficient to solely rely on this short period of history. Another interesting fact is that the model seems to perform worse in predicting the motions of the right elbow joint. A possible reason could be that the movements of the right elbow are also affected by the pose of the pelvis, while the knees have a more independent thus also more predictable motion pattern.

\subsubsection{Qualitative Evaluation}
\label{sec: qualitative evaluation}
\begin{figure*}[h]
    \centering
    \includegraphics[width=1.0\textwidth]{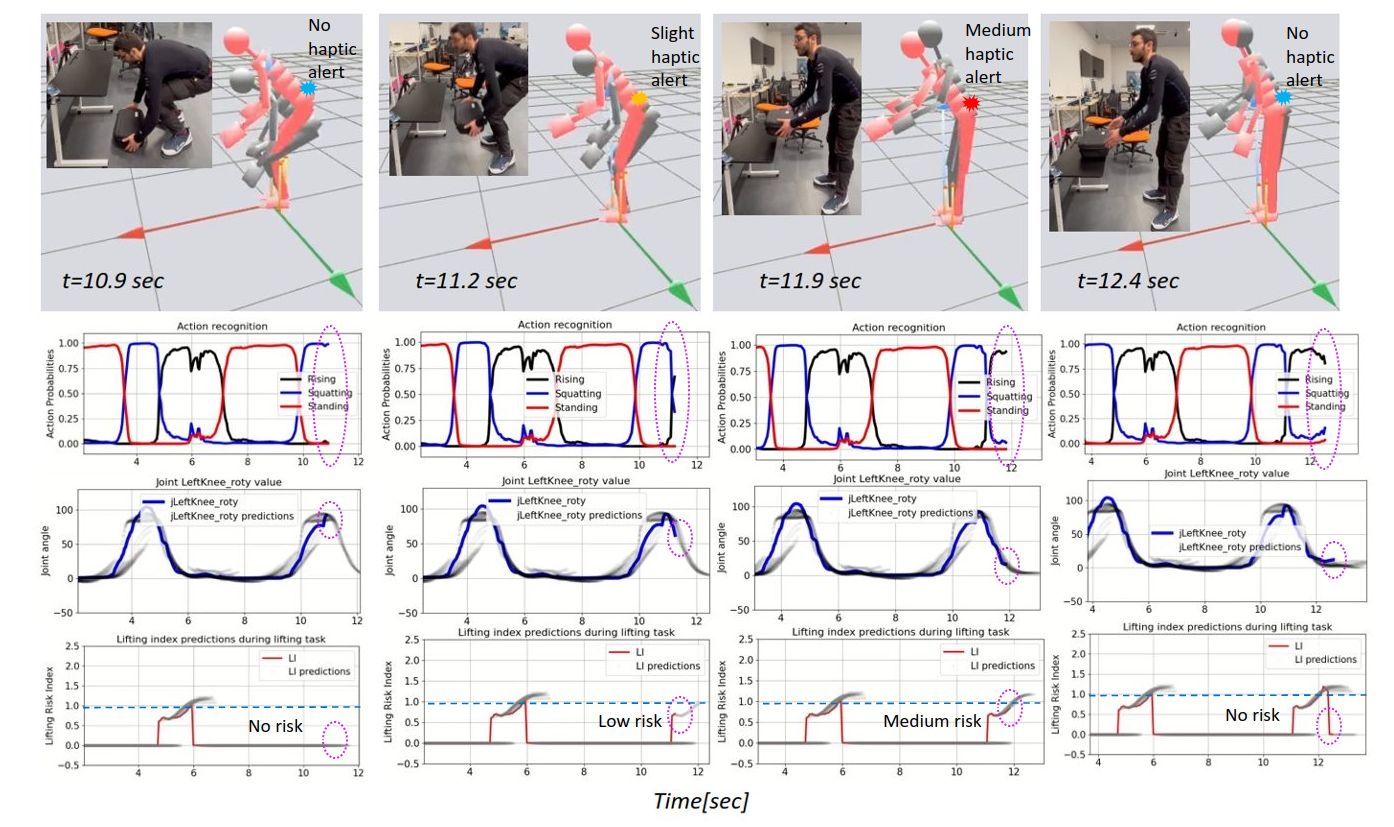}
    \caption{Experimental results of online action recognition and risk prediction. The first row shows pictures of the sensorized subject during the task execution and virtual model visualization with estimated (gray) and predicted (red) configuration. In the second row, it is shown the action prediction probability. In the third row, ground truth and prediction of the left knee joint rotation angle are depicted. The bottom row shoes lifting index for the period till prediction time horizon.}
    \label{fig:resutls}
\end{figure*}

To further evaluate the effectiveness of the proposed framework, we analyze qualitatively the results of \emph{Task 2} (shown in Table \ref{tab:task_data}) as an example. A complete process of \emph{rising} is demonstrated in Figure \ref{fig:resutls}. As shown in the first row, the motions of both real human subject and simulated models are captured. The grey model reconstructs the human motion at current time \emph{t} from sensor measurements, while the red model represents the predicted human motion at future time \emph{t+0.6s} (in the experiment we output the maximum future 20 data frames, recalling the period between each data frame is 30ms, thus the prediction time is 0.6s). The correspondingly recognized actions at each moment are presented in the second row. The black, blue and red solid curves denote the probability of action \emph{rising}, \emph{squatting} and \emph{standing}, respectively. In the third row, we demonstrate the predictions of rotation degrees of left knee joint around \emph{y}-axis for future 1.5s (maximum 50 future data frames), associated with the round dot curves. In the meanwhile, the blue curve stands for the ground truth of left knee joint rotation values. Figures in the last row demonstrate the lifting index during the \emph{rising} action. The red curve and grey dot curve represent the risk value at the current time and future 0.9s (namely 30 data frames), respectively.

As shown in the picture at the top left in Figure \ref{fig:resutls}, the human is almost finishing the action \emph{squatting} at t=10.9s, and as the red model indicates, at the future time t=11.5s, the human model would probably be rising up a little bit. The recognized action at t=10.9s is still \emph{squatting}, therefore no lifting risk is detected and the haptic actuator remains silent. As for the rotation angle of the left knee joint, it also reaches a peak value of about 100 degrees and it's going to decrease soon. When time t becomes 11.2s, it can be observed that the gray model is reaching the pose as predicted at t=10.9s. In the meanwhile, the action transition already started, thus we can see that the lifting risk grows from zero to 0.7 (hence a slight haptic alert appears), and as the predictions show, the risk value at t=11.8s should be equal to 1.0. Then at t=11.9s, the human is reaching the table and intends to put the payload on it. At this moment, the action is still recognized as \emph{rising} with maximum probability. Moreover, the currently estimated lifting index is around 0.9 (corresponding to a medium haptic warning), which almost equals the value predicted at t=11.2s. At the final time t=12.4, apparently the \emph{rising} action is completed, and the human subject is getting back to \emph{standing} pose. Therefore the probability of \emph{rising} starts to decrease. Correspondingly, the lifting index returns back to zero again.


\subsubsection{Failure Cases}
We present some failure cases here to reveal the limitations of the current system. As explained in Section \ref{subsec:make_data}, the GMoE network is trained on a 15-mins data set that consists of basic lifting tasks. Hence, a very typical unsuccessful scenario is when completely unseen motion patterns appear in the online application, e.g., trunk twisting and overhead lifting. In such cases, precise action detection can become an issue, let alone predict risks. Additionally, the GMoE model can be further generalized when trained on a dataset with multiple individuals (e.g., age, gender, body shape and etc). Another challenge lies in the restrictions of the NIOSH equation.  For example, the system is not applicable to collaborative lifting tasks where multiple workers are present. Moreover, the noise and perturbations accumulated over time in online applications also have a great effect on the accuracy of the GMoE model. We hypothesize that the retrievement of unprecise NIOSH variables is also a notable limitation. This is the main reason for improving the swiftness of action detection and the accuracy of motion predictions.

\subsection{Discussion}
In comparison to risk assessment approaches proposed in literature \cite{Shafti2019, Fortini2020ROMAN, fortini2020framework}, the main advantage of the proposed framework lies in its ability to early assessment and prevention of biomechanical risks faced by workers during realistic lifting tasks, by utilizing a learning-based approach and wearable sensing system. Despite training on a relatively small data set, we have shown that our model is able to generalize well to unseen data (though restricted to the same motion patterns that appeared in the training dataset), as analyzed in \ref{sec: analysis action recognition} and \ref{sec: analysis motion prediction}. We also demonstrate robust qualitative performance during the live demo presented in \ref{sec: qualitative evaluation}. It is worth mentioning that although humans can feel muscular fatigue in the long term, the causal action is often neglected due to the lack of real-time quantitative ergonomic feedback. Therefore the anticipated haptic alerts is very potentially to improve the risk awareness of workers while performing heavy lifting tasks.

%% file: sections/05_conclusions.tex
\section{CONCLUSIONS}
\label{sec:conclusions}
In this paper, we presented a framework that integrates wearable sensing, human state estimation, human action/motion prediction and NIOSH index for real-time manual lifting applications. Through online recognition of human actions, the execution of a single lifting activity can be segmented into a series of continuous parts. The commencement of each sub-action is considered the initial human state, with subsequent moments within this sub-action being regarded as temporary destination states. With the help of motion prediction, future human status can also be obtained. Hence RNLE can be applied to assess risks within the predicted time horizon. The vibrotactile feedback enables anticipated alert on the predicted lifting risks. The performance of the framework is tested in an experimental lifting scenario using the iFeel wearable system.

Future work should first address the problem of generalization by expanding the current lifting data set, such that more complex realistic lifting tasks can be considered. By improving the performance of GMoE model, a more precise retrieval of NIOSH geometry variables could be expected. It would also be interesting to include upper trunk twisting and overhead lifting in order to utilize the NIOSH equation. Moreover, a learning-based ergonomics assessment approach could be another promising topic.  